\documentclass[pra,twocolumn,superscriptaddress,showpacs,preprintnumbers,amsmath,amssymb]{revtex4-1}
\usepackage{graphicx}% Include figure files
\usepackage{dcolumn}% Align table columns on decimal point
\usepackage{bm}% bold math
\usepackage{multirow}
\usepackage{color}
\usepackage{ifpdf}
\ifpdf
	\usepackage{epstopdf}
	\pdfpageattr {/Group << /S /Transparency /I true /CS /DeviceRGB>>}
\fi

\begin{document}

\bibliographystyle{apsrev4-1} % Choose Phys. Rev. style for bibliography, Rev.4

%\preprint{APS/123-MW}

\title{Coherent zero-field magnetization resonance in a dipolar spin-1 Bose-Einstein condensate}

\author{Wenxian Zhang}
\affiliation{School of Physics and Technology, Wuhan University, Wuhan, Hubei 430072, China}
\affiliation{Beijing Computational Science Research Center, Beijing 100084, China}

\author{S. Yi}
\affiliation{Institute of Theoretical Physics, Chinese Academy of Sciences, Beijing 100190, China}

\author{M. S. Chapman}
\affiliation{School of Physics, Georgia Institute of Technology, Atlanta, Georgia 30332-0430, USA}

\author{J. Q. You}
\affiliation{Beijing Computational Science Research Center, Beijing 100084, China}

\date{\today}

\begin{abstract}
With current magnetic field shielding and high precision detection in dipolar spinor Bose-Einstein condensates, it is possible to experimentally detect the low or zero field nonsecular dipolar dynamics. Here we analytically investigate the zero-field nonsecular magnetic dipolar interaction effect, with an emphasis on magnetization dynamics in a spin-1 Bose-Einstein condensate under the single spatial mode approximation within the mean field theory. Due to the biaxial nature of the dipolar interaction, a novel resonance occurs in the condensate magnetization oscillation, contrast to the previous assumption of a conserved magnetization in strong magnetic fields. Furthermore, we propose a dynamical-decoupling detection method for such a resonance, which cancels the stray magnetic fields in experiments but restores the magnetization dynamics. Our results shed new lights on the dipolar systems and may find potential applications beyond cold atoms.
\end{abstract}

% PACS, the Physics and Astronomy Classification Scheme.
% 03.75.Kk Dynamic properties of condensates; collective and hydrodynamic
%          excitations, superfluid flow
% 03.75.Mn Multicomponent condensates; spinor condensates
% 05.30.Jp Boson systems (for static and dynamic properties of Bose-Einstein
%          condensates, see 03.75.Hh and 03.75.Kk)
% 42.65.-k Nonlinear optics
% 05.45.Gg Control of chaos, applications of chaos
% 67.85.Fg Multicomponent condensates; spinor condensates

\maketitle

\section{Introduction}
Magnetic dipolar interactions exist in a wide variety of physical systems, including nuclear spins~\cite{Slichter92}, atoms and molecules~\cite{Baranov08, Lahaye09}, condensed matter~\cite{Teixeira00}, chemical~\cite{Prestegard04} and biological systems~\cite{Blackledge05}. Due to the relative weakness of the magnetic dipolar interaction to the Zeeman effect of an external magnetic field, the nonsecular interaction in a dipolar system, which varies temporally with Larmor frequency and breaks the rotational invariance, was usually neglected in most experiments over decades. Such an approximation results in the famous magnetization conservation during the time evolution~\cite{Slichter92}.

Recently, with the precision measurement of an ultraweak or zero magnetic field, the investigation of the nonsecular dipolar interaction  effects becomes revived in the areas of zero-field nuclear magnetic resonance and dipolar spinor Bose-Einstein condensates~\cite{Ledbetter11, Theis11, Sheng13, Pasquiou11, Eto13}. Particularly, the achievements in spinor Bose-Einstein condensates (BECs) provide a highly tunable and controllable system where the spin interactions, including the magnetic dipolar interaction, can be accurately engineered~\cite{Lahaye09, Stamper-Kurn13, Yi04, Cheng05, Widera05, Sun06, Yi06, Kawaguchi06, Kawaguchi10}. Such dipolar spinor condensates make them an ideal testbed to extensively explore the full magnetic dipolar interaction effect, including not only the secular part but also the nonsecular one which breaks the rotational symmetry and definitely brings new physics. Recent experiments by Pasquiou {\it et al.} investigated the incoherent demagnetization process in spin-3 $^{52}$Cr condensates in an ultralow magnetic field~\cite{Pasquiou11}. However, the coherent dynamics induced by the nonsecular dipolar terms in cold atoms has yet been touched.

In this paper, we investigate the effect of the magnetic dipolar interaction in a spin-1 BEC, focusing on the coherent magnetization dynamics induced by the nonsecular dipolar interaction in low or zero field (See Fig.~\ref{fig:t}). By adopting mean field theory and single spatial mode approximation (SMA)~\cite{Pu99, Yi02, Zhang03}, we analytically obtain the magnetization dynamics of the spin-1 BEC. Contrast to the usual assumption of magnetization conservation, we find a novel resonance in the magnetization oscillation in the dipolar condensate. With large-scale numerical calculations, we also explore the spin dynamics beyond the SMA, which agrees well with the analytical prediction. Furthermore, a practical control protocol utilizing dynamical decoupling techniques is proposed to liberate the nonsecular part of the magnetic dipolar interaction by canceling the stray magnetic fields. Our results shed new light on the dipolar systems, in particular on the resonant magnetization dynamics induced by the nonsecular part of the dipolar interaction, and open a door to understanding the complex spin texture observed in dipolar spinor BECs~\cite{Chang05, Sadler06, Cherng09, Kronjager10, Eto14}.

The paper is organized as follows. In Sec.~\ref{sec:ham}, we introduce the mean field Hamiltonian describing the dipolar spin-1 condensate at zero magnetic field. In Sec.~\ref{sec:ana}, we present the equation of motion for the condensate magnetization and the analytical solution. We then design dynamical decoupling pulse sequence to suppress the Zeeman effect of a nonzero magnetic field and perform numerical simulation with parameters in practical experiments beyond SMA in Secs.~\ref{sec:dd} and \ref{sec:num}, respectively. We conclude in Sec.~\ref{sec:con}.

\section{System Hamiltonian}
\label{sec:ham}
In the mean field theory and under the SMA, a dipolar spin-1 Bose condensate is described by the Hamiltonian~\cite{Yi04, Ning12, Huang12},
$$
    H = c f^2 + d_s (3f_z^2-f^2)+3d_n (f_y^2-f_x^2),
$$
where the spin exchange interaction strength is $c=(c_2/2)\int d\mathbf{r}\rho^2(\mathbf r)$ with $\rho(\mathbf r) = |\phi(\mathbf r)|^2$ being the density of the mode function $\phi(\mathbf r)$. The spin-exchange interaction coefficient $c_2 = 4\pi \hbar^2(a_2-a_0)/(3M)$ with $M$ being the atom's mass and $a_{0,2}$ the $s$-wave scattering length of two spin-1 atoms in the symmetric channel of the total spin $0$ and $2$, respectively. The condensate spin $f^2 = f_x^2 + f_y^2 + f_z^2$, where $f_{x,y,z} = \langle {\vec \xi} | F_{x,y,z} |{\vec \xi} \rangle$ with $|\vec \xi\rangle = (\xi_+, \xi_0, \xi_-)^T$ being the spin wave function and $F_{x,y,z}$ the spin-1 matrices. Note that we have neglected the constant terms under the SMA, contributed by the kinetic energy, the trapping potential, and the spin-independent interaction proportional to $c_0 = 4\pi \hbar^2(2 a_2+a_0)/(3M)$.

The dipolar interaction includes two parts, the secular part with $d_s = (c_d/4)\int d\mathbf{r}d\mathbf{r'}{|\mathbf{r}-\mathbf{r'}|^{-3}} \rho(\mathbf r)\rho(\mathbf{r'})(1-3\cos^2\theta_e)$ and the nonsecular part with  $d_n=(c_d/4)\int d\mathbf{r}d\mathbf{r'} {|\mathbf{r}-\mathbf{r'}|^{-3}} \rho(\mathbf{r})\rho(\mathbf{r'})\sin^2\theta_e e^{i2\varphi_e}$, where $\theta_e$ and $\varphi_e$ are respectively the polar and azimuthal angles of $(\mathbf{r}-\mathbf{r}')$. The dipolar interaction coefficient is $c_d = \mu_0\mu_B^2 g_F^2/(4\pi)$ with $\mu_0$ being the vacuum permeability, $\mu_B$ the Bohr magneton, and $g_F$ the Land\'e $g$-factor for an electron. For $^{87}$Rb atoms, $c_d/|c_2| \approx 0.09$, and for $^{23}$Na atoms, $c_d/|c_2| \approx 0.007$. Note that $d_n = 0$ if $\rho(\mathbf r)$ is axially symmetric. In general, the value of $d_{s,n}$ can be positive, 0, or negative, depending on the specific shape of the mode function~\cite{Huang12, ntdsn}. In a large or moderate magnetic field, the secular part survives due to the conservation of $f_z$, but the nonsecular part, which flips the spin, is strongly suppressed by the Zeeman effect and usually neglected. In Sec.~\ref{sec:dd}, we will propose a dynamical decoupling method to cancel the Zeeman effect and to ``restore" the suppressed nonsecular part.

The original Hamiltonian can be rewritten as
\begin{equation}
    H = A_0 f^2 + A_x f_x^2 + A_z f_z^2
\end{equation}
where $A_0 = c-d_s+3d_n$, $A_x = -6d_n$, and $A_z = 3(d_s-d_n)$. We have used the relation that $f_y^2 = f^2-f_x^2-f_z^2$. For such a spin system, besides the conservation of the total energy and the spin $f$ of the condensate (thus the conservation of the isotropic energy $E_0 = A_0 f^2$), we notice that the sum of the rest two anisotropic terms, which we define as $E_{xz} = A_x f_x^2 + A_z f_z^2$, is also conserved. The conservation of $E_{xz}$ defines the spin trajectories in the $x$-$z$ plane into the following two categories, depending on the values of $A_x$ and $A_z$ (besides the trivial one, $A_x A_z = 0$ where $f_z$ or $f_x$ is constant): (I) $A_x A_z > 0$, the trajectory in the $x$-$z$ plane is an ellipse (or a circle for $A_x = A_z$); (II) $A_x A_z < 0$, the trajectory is a hyperbola (or a parabola for $A_x = -A_z$). Of course, $f_{x,z}$ must be smaller than ($f_y \neq 0$) or equal to $f$ ($f_y=0$).

\begin{figure}
  \includegraphics[width=3in]{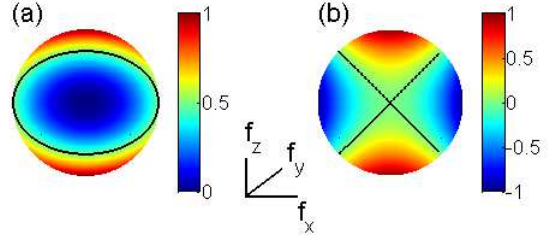}
  \caption{(Color online) The iso-energy contour plot of the anisotropic energy $E_{xz}$ in unit of $|c|$ on the surface of a spin sphere (a) for category (I) with $A_x = 0.5$ and $A_z = 1$ and (b) for category (II) with $A_x = -1$ and $A_z = 1$. The black lines mark the boundary between the $z$ and $y$ regions in (a) and between the $z$ and $x$ regions in (b).}
  \label{fig:exz}
\end{figure}

The iso-energy contour plots of $E_{xz}$ on the surface of the spin sphere for categories (I) and (II) are presented in Fig.~\ref{fig:exz}. For both categories the oscillations of $f_z$ either passes zero or not, depending on the initial value of $f_z$. As shown in the figure, there is an alternative way to classify the spin dynamics: the $x$-, $y$-, or $z$-region, where the spins rotates around the corresponding $x$-, $y$-, or $z$- axis. Actually, the trajectory of the spin behaves like nothing but a nonlinear rigid pendulum, which oscillates around $x$-, $y$-, or $z$- axis correspondingly.

\section{Analytical results of magnetization resonance under SMA}
\label{sec:ana}
By treating the condensate spin as a classical spin, which rotates in an effective magnetic field $(b_x, b_y, b_z) = (2A_x f_x, 0, 2A_z f_z)$, we obtain the following equation of motion for the magnetization,
$
    \dot f_z \equiv {d f_z}/{d t} = 2 A_x f_x f_y.
$
By utilizing again the relations $f_y^2 = f^2 - f_x^2 - f_z^2$ and $f_x^2 = (E_{xz} - A_z f_z^2)/A_x$, we find a closed equation of motion for $f_z$,
\begin{equation}\label{eq:fz}
    {\dot f_z}^2 = 4(A_zf_z^2 -E_{xz})\left[(A_x-A_z)f_z^2 + E_{xz} - A_x f^2\right],
\end{equation}
where $f$ and $E_{xz}$ are determined by the initial state.

The analytical solution for $f_z$ is an inverse function of the elliptic integral of the first kind $F[\cdot, \cdot]$,
\begin{equation}\label{eq:sol}
    t = t_0 + \frac{1} {2\sqrt{A_z (A_x f^2-E_{xz})}} \, F\left[\sin^{-1}\left(\frac{f_z}x\right), \frac{x^2}{y^2}\right],
\end{equation}
where $x^2 = E_{xz}/A_z$ and $y^2 = (A_xf^2-E_{xz})/(A_x-A_z)$. The value of $x$ ($y$) is determined by $f_{x(y)} = 0$ (i.e., $\dot f_z =0$).

The period of the oscillation is
\begin{eqnarray}\label{eqn:prd}
    T &\equiv & \oint \dot f_z(t)^{-1} df_z \nonumber \\
    &=& \frac{2} {\sqrt{A_z (A_x f^2-E_{xz})}} \, \left|F\left[\sin^{-1}\left({y\over x}\right),\frac{x^2}{y^2}\right]\right|
\end{eqnarray}
if the oscillation is in the $x$-region,
\begin{eqnarray}
    T &=& \frac{2} {\sqrt{A_z (A_x f^2-E_{xz})}} \, K\left(\frac{x^2}{y^2}\right) \nonumber
\end{eqnarray}
if in the $y$-region, or
\begin{eqnarray}
    T &=& \frac{2} {\sqrt{A_z (A_x f^2-E_{xz})}} \, \left|F\left[\sin^{-1} \left(\frac y x\right), \frac{x^2}{y^2}\right] - K\left(\frac{x^2}{y^2}\right) \right| \nonumber
\end{eqnarray}
if in the $z$-region, where $K(\cdot)$ is the complete elliptic integral of the first kind. The function $K(k)$ is nearly a constant $\pi/2$ around $k=0$ and diverges rapidly if $k\rightarrow 1$. In addition, $K(k) = F(\pi/2,k)$, and $K(1/k)/k^{1/2}$ also diverges if $k\rightarrow 0$.

\begin{figure}
  \includegraphics[width=3in]{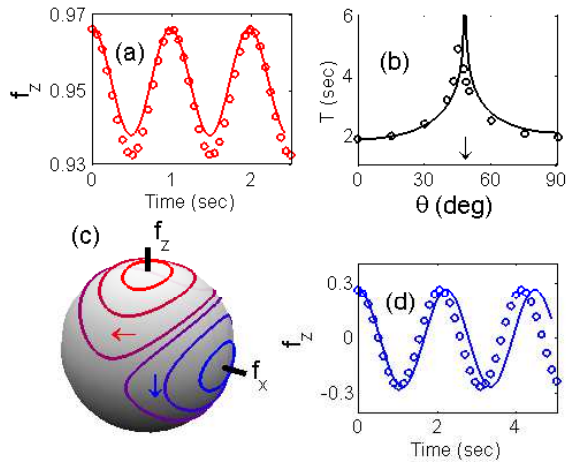}
  \caption{(Color online) Periodic spin dynamics induced by the magnetic dipolar interaction. Oscillation of the magnetization $f_z$ with an initial rotation angle (a) $\theta = 15^o$ (below the critical value $\theta_c$) and (d) $\theta = 75^o$ (above $\theta_c$), where the spin oscillates around $z$- and $x$-axis, respectively. (b) Dependence of the oscillation period $T$ on the initial angle $\theta$. Solid lines are the SMA prediction and circles are the numerical results with coupled Gross-Pitaevskii equations. A resonance occurs at $\theta_c$, marked by the black arrow in (b), where the spin evolves along neither $x$- nor $z$- axis. (c) Typical iso-energy spin trajectories of the condensate on the surface of the spin sphere. The red and blue arrows show the spin evolution direction.}
  \label{fig:t}
\end{figure}

The analytical solution for $f_z$ shows oscillatory and periodic motion. This is in sharp contrast to the previous understanding of the conservation of the magnetization (in large magnetic fields), because the nonsecular terms break the rotational symmetry around $z$-axis. As shown in Fig.~\ref{fig:exz}, the amplitude of the oscillation of $f_z$ is $|y|$ if the condensate spin is in the $x$-region, $|x|$ if in the $y$-region, or $|x-y|$ if in the $z$-region. Remarkably, the period $T$ diverges if $x=0$ ($y=0$), i.e., the initial spin state is set on the boundary between the $x$-region ($y$-region) and the $z$-region (see Fig.~\ref{fig:exz}). This divergence indicates that a resonance occurs by changing the initial spin state across the boundary. Such an interesting resonance has not been revealed before, because of the neglect of the nonsecular terms, and is obviously due to the competition between the secular and nonsecular dipolar terms.

For a clear view, we present in Fig.~\ref{fig:t} the analytical results for the category (II) with $d_s/|c| = 0.225$ and $d_n/|c| = 0.0644$ ($|c| = 1.3$ Hz is calculated for a numerically simulated spin-1 BEC). Two typical oscillations of $f_z$ in the $z$-region and $x$-region are shown in Fig.~\ref{fig:t}(a) and (d). Other trajectories are shown in Fig.~\ref{fig:t}(c) and the periods of the oscillations with respect to initial polar angle $\theta$ ($f=1$ and zero azimuthal angle) of the condensate spin are shown in Fig.~\ref{fig:t}(b). We observe a clear resonance signature at $\theta_c$, which is determined by $E_{xz} = 0$, i.e., $\cos^2 \theta_c = A_x / (A_x - A_z)$. Similarly, for the category (I) we also observe a resonance in the $f_z$ oscillation period and the critical angle $\cos^2 \theta_c = A_x / A_z$, which is obtained from $A_x f^2 - E_{xz} = 0$~\cite{diplifetime}.

The resonance in the oscillation period can be understood physically by treating the dipolar spin-1 BEC as a two-axis rigid nonlinear pendulum (see Fig.~\ref{fig:t}). For a single-axis pendulum, i.e., either $A_x=0$ or $A_z=0$, the condensate spin oscillates around an axis, $z$ or $x$, with a constant spin component along this axis. But in categories (I) and (II), neither of $A_{x,z}$ is zero, so the evolution of the condensate spin is not purely around a single axis but a more complicated oscillation between the two axes, due to the competition of the two nonlinear terms of $A_x f_x^2$ and $A_z f_z^2$. In general, by setting the condensate spin initially closer to an axis, e.g., $z$-axis, the evolution of the condensate spin is around this axis since the corresponding term ($A_z f_z^2$) is dominant. However, there exists a clear boundary where the two-axis terms are balanced, e.g, $A_x f_x^2 = -A_z f_z^2$ in Fig.~\ref{fig:exz}(b), and the condensate spin evolves in a third direction and reaches to a dead end ($f_{x,z} = 0$). The oscillation never completes and the period becomes infinite. Thus a resonance peak appears in the spin oscillation period.

An alternative way of understanding the resonance behavior is to treat the condensate in a semiclassical manner. In a single particle picture, the energy level of an atom in the $|m_F= \pm 1\rangle$ state is shifted by an amount of $A_z$ due to the term $A_z f_z^2$. Similarly, the term $A_x f_x^2$ not only shifts oppositely the energy level of $|m_F = \pm 1\rangle$ (if $A_xA_z<0$), but also couples them. As shown in Fig.~\ref{fig:t} and~\ref{fig:exz}(b), when the resonant condition $A_x f_x^2 + A_z f_z^2 = 0$ is satisfied and the condensate reaches its steady state ($f_{x,z} = 0$), the total shift of the energy levels of $|m_F = \pm 1\rangle$ become zero, i.e., the energy levels of $|m_F=\pm 1\rangle$ are resonant to that of $|m_F = 0\rangle$.

\begin{figure}
  \includegraphics[width=3in]{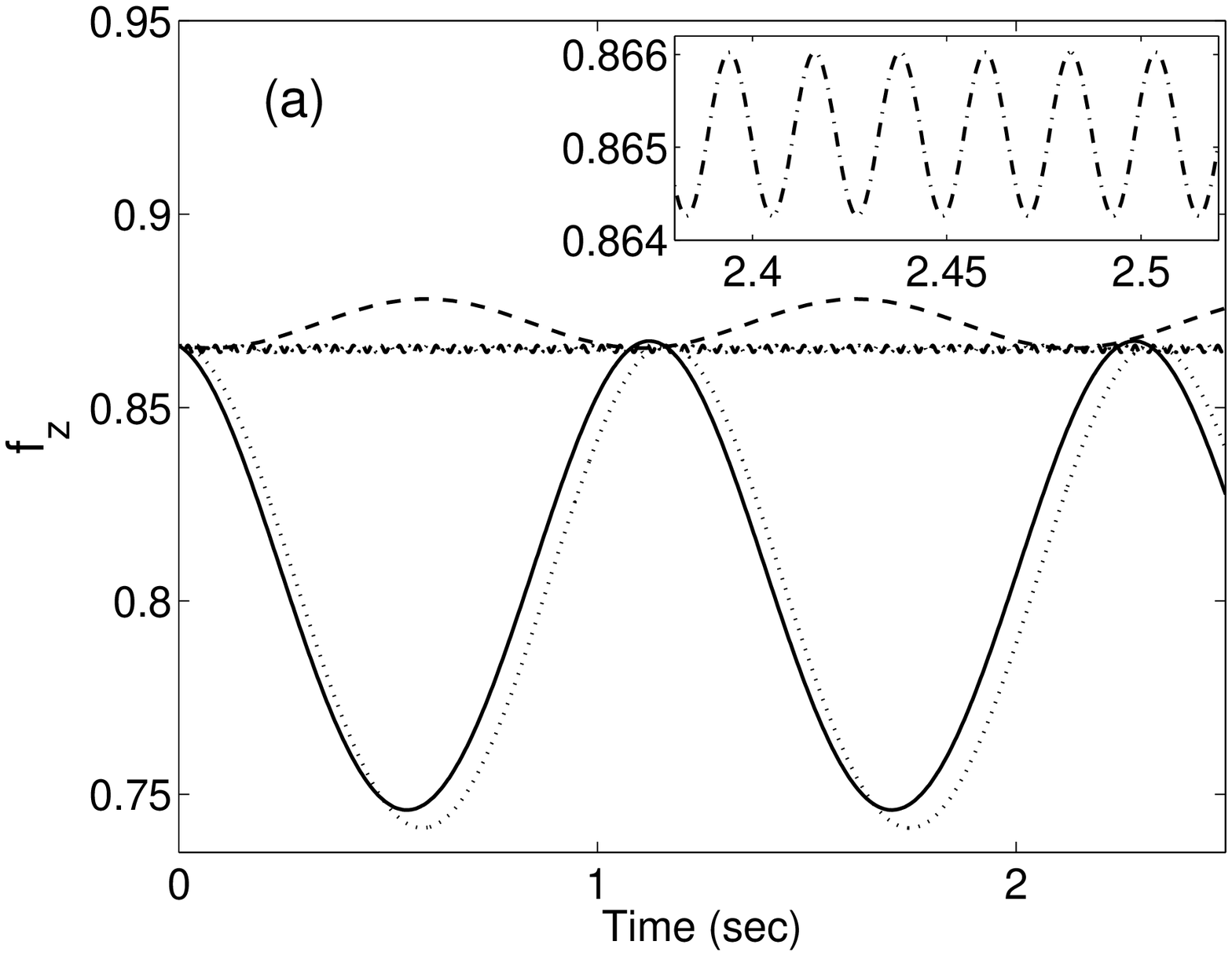}
  \includegraphics[width=3in]{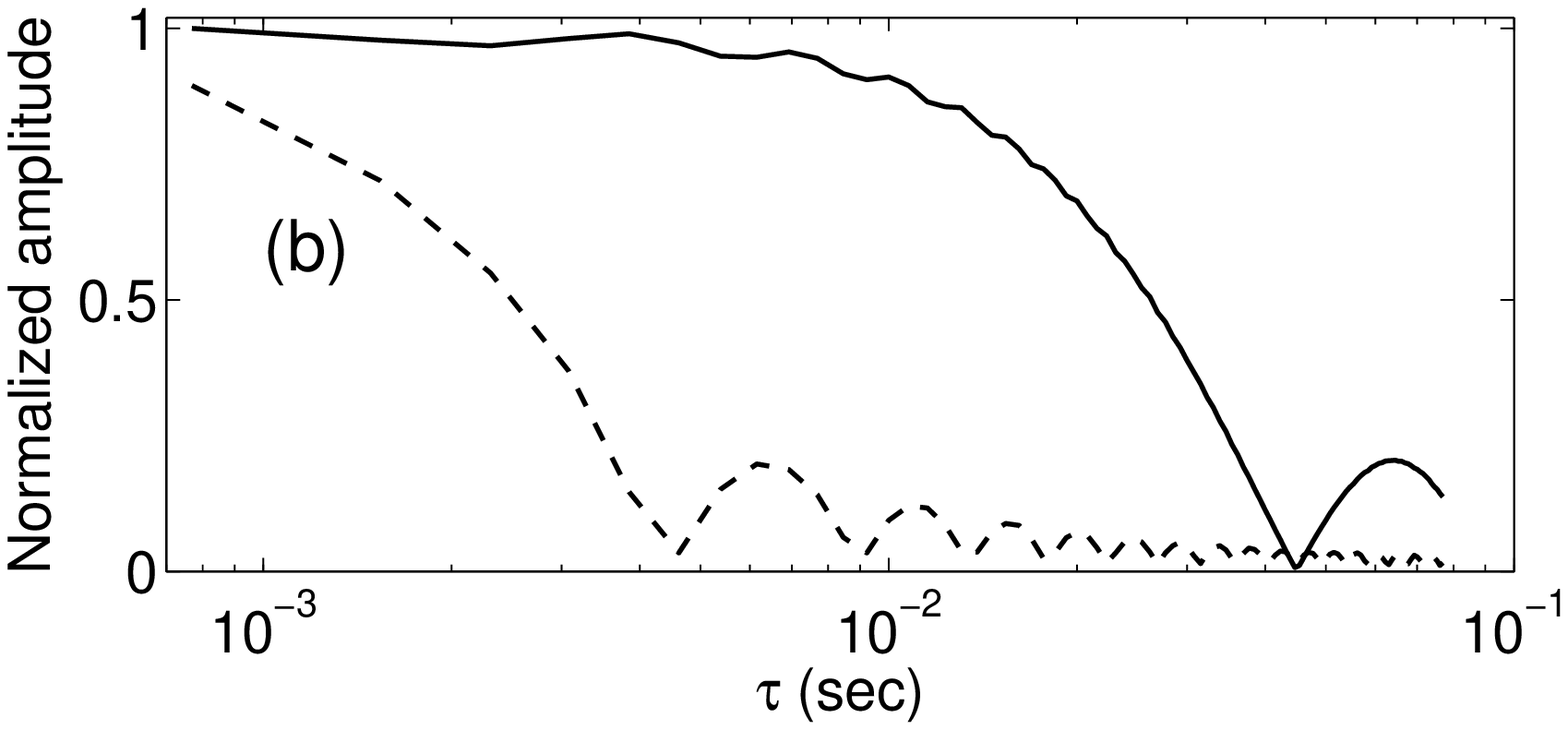}
  \caption{(a) Coherent magnetization dynamics of a dipolar spin-1 BEC in a magnetic field. $B=0$ (dotted line), $B=0.1$ mG (dash-dotted line), $B=0.1$ mG with control pulse delay $\tau = 0.004$ s (solid line), and $B=0.1$ mG with $\tau = 0.04$ s (dashed line). The initial condition is the same as in Fig.~\ref{fig:t} with $\theta = 30^o$. Other parameters are the same as in Fig.~\ref{fig:t}. The Zeeman effect of the nonzero magnetic field suppresses the $f_z$ dynamics (dash-dotted line and the inset for a zoom-in view), but the fast and periodic application of $\pi_x$ pulses restores the dipolar $f_z$ dynamics. (b) Dependence of the normalized amplitude of the dipolar magnetization oscillations on the control pulse delay $\tau$ at $B=0.1$ mG (solid line) and $B=1$ mG (dashed line).}
  \label{fig:b}
\end{figure}

\section{Suppressing the Zeeman effect of an external strong magnetic field}
\label{sec:dd}
Inside a magnetic shield room, the stray magnetic field have been reduced to as low as 0.1 mG~\cite{Pasquiou11, Eto13}. However, even in such a low magnetic field, the Zeeman effect still overwhelms the dipolar-interaction-induced spin dynamics along the $z$-axis, since the Zeeman energy $\mu_B B \sim 140$ Hz for $B=0.1$ mG is much larger than the dipolar interaction energy, typically in the order of $0.1$ Hz at a condensate density of $10^{14}$ cm$^{-3}$. As shown in Figs.~\ref{fig:b}(a), the oscillation amplitude of $f_z$ for $B=0.1$ mG (dash-dotted line and the inset) is much smaller than that for $B=0$ (dotted line). The consequence is that the magnetization dynamics due to the dipolar interaction is rather challenging to observe in a practical experiment because the stray magnetic field effects dominate the magnetization dynamics.

However, the resonant dipolar magnetization dynamics can be revealed under current experimental conditions if we employ dynamical decoupling techniques to cancel the Zeeman effect~\cite{Slichter92, Ning11, Ning12}. In particular, by applying frequent $\pi_x$ pulse, which rotates the condensate spin 180$^o$ along $x$-axis, the Zeeman effect is eliminated while leaving the dipolar interaction intact to the leading order. We assume that the control pulse $\pi_x$ is an instantaneous (hard) pulse and the delay between two adjacent pulses is $\tau$~(Appendix~\ref{appd:dd}). The numerical simulation results are presented in Fig.~\ref{fig:b}. As shown in Fig.~\ref{fig:b}(a), in the limit of small $\tau$ (solid line), the dipolar magnetization dynamics is restored; in the limit of large $\tau$ (dashed line), the Zeeman effect is not well suppressed and the dipolar magnetization oscillation amplitude is small. Under control pulses, the dependence of the normalized magnetization oscillation (which is the ratio of the  magnetization oscillation amplitude at $B=0.1$ mG under control pulses to the free magnetization oscillation at $B=0$) on the pulse delay $\tau$ is shown in Fig.~\ref{fig:b}(b). Obviously, the Zeeman effect is canceled and the magnetization oscillation is well restored at $B=0.1$ mG if $\tau$ is smaller than $0.02$ seconds, which is easily realizable in experiments~(Appendix~\ref{appd:dd}).

\section{Numerical results with a large number of atoms}
\label{sec:num}
Our previous analysis are based on the SMA, which may not be valid in some experiments for spin-1 BECs, particularly for a condensate with a large number of atoms~\cite{Vengalattore08, Yi06, Kawaguchi07}. It is an open question whether the SMA results remain valid under practical experimental conditions, where the number of $^{87}$Rb atoms is $10^4$ and the trap frequencies are $\{\omega_x, \omega_y, \omega_z\} = (2\pi \times) \{90, 140, 200\}$ Hz. For this trap geometry, the SMA with a three-dimension Gaussian wave function predicts that $d_s = 0.215$ and $d_n = 0.066$, while the numerical results for a fully polarized ground state (all atoms are in $|+1\rangle$ state) are $d_s = 0.225$ and $d_n = 0.064$. It shows obviously that the SMA is slightly violated.

We start the numerical simulation by tilting the fully polarized ground state away from the $z$-axis by a polar angle $\theta$ in the $x$-$z$ plane. In this way, the subsequent magnetization evolution is solely due to the dipolar interaction since the condensate magnetization would remain constant in the absence of the anisotropic dipolar interaction. The magnetization dynamics of the condensate is obtained by numerically solving the three coupled Gross-Pitaevskii equations in the given trap\cite{Zhang09, Kawaguchi07, Ning12}
\begin{eqnarray}
\label{eq:gp}
i\hbar\frac{\partial\psi_{\alpha}}{\partial t}&=&\left[T+V_{\rm ext}+c_0 n\right]\psi_\alpha + {\mathbf B}_{\rm eff} \cdot {\mathbf F}_{\alpha\beta} \psi_\beta,
\end{eqnarray}
where the kinetic energy is $T = -\hbar^2\nabla^2/(2M)$ with $M$ being the atom mass of $^{87}$Rb, the trapping potential is $V_{\rm ext}(x,y,z) = M(\omega_x^2 x^2 +\omega_y^2 y^2 +\omega_z^2 z^2)/2$, and the total number density is $n=\sum_\alpha \psi^*_\alpha\psi_\alpha$ with $\psi_{\alpha}$ ($\alpha=\pm1,0$) being the three components of the condensate wave function. The effective field originating from the spin-exchange and dipolar interactions is
$
{\mathbf B}_{\rm eff} = c_2 {\mathbf S} + c_d \int d{\mathbf r}' \{{\mathbf S}({\mathbf r}')-3[{\mathbf S}({\mathbf r}')\cdot {\mathbf e}]{\mathbf e}\} / {|{\mathbf r}-{\mathbf r}'|^3},
$
where $\mathbf{S}= \psi^*_\alpha \mathbf{F}_{\alpha\beta} \psi_\beta$ is the spin density with $\mathbf{F}$ the atom spin-1 matrix, and $\mathbf{e}$ is the unit vector along $\mathbf{r}-\mathbf{r}'$. The spin-exchange interaction coefficient $c_2$ and the dipolar interaction coefficient $c_d$ are given previously. We solve these coupled equations using the operator splitting method, where the term involving the integral operator $\mathbf B_{\rm eff}$ is calculated with convolution theorem and fast Fourier transform.

We present in Figs.~\ref{fig:t}(a), \ref{fig:t}(b), and \ref{fig:t}(d) the numerical results, as well as the SMA predictions with $d_{s,n}$ obtained numerically from the fully polarized ground state. We observe a pretty good agreement of the spin oscillations between the numerical results and the SMA prediction, due to the fact that the three components of the ground state of the ferromagnetically interacting spin-1 condensate share the same spatial wave function~\cite{Yi02, Zhang03}. However, we find slight mismatches of the oscillation amplitude in Fig.~\ref{fig:t}(a) and the resonance peak position in Fig.~\ref{fig:t}(b). These discrepancies might be due to the trap anisotropy: the atoms spins are more likely to align along the loosely trapped $x$- or $y$-axis so that the total dipolar energy is lower, which means that the $z$-region is smaller than the SMA results. Consequently, the boundary between the $x$-region and the $z$-region is shifted to a smaller $\theta_c$, which is what we observe in Fig.~\ref{fig:t}(b), and the lower value of $f_z$ in Fig.~\ref{fig:t}(a) becomes smaller.

\section{Conclusion}
\label{sec:con}
The weak dipolar interaction effect, particularly the effect of the nonsecular part, in spin-1 Bose condensates such as $^{87}$Rb is experimentally challenging to observe~\cite{Kawaguchi07, Gawryluk11}. Our proposal provides a practical way to detect the condensate magnetization oscillation induced by the nonsecular dipolar interaction, where a resonance emerges in the oscillation period. Numerical results with experimental parameters, such as the large number of atoms and the suppression of Zeeman effect of the stray magnetic field with dynamical decoupling method, confirm that the resonant behavior of the coherent magnetization dynamics is detectable experimentally. Our results point to a new direction for future investigations on many dipolar effects, including the competition with the short-range contact interaction, the quantum dipolar effects which are of great interests for quantum metrology and next-generation-precision magnetometers based on spinor Bose condensates~\cite{Bookjans11, Hamley12, Gerving12, Hoang13, Vengalattore07, Eto13}, and the structure determination in chemical and biological dipolar systems.

\acknowledgements
W.Z. thanks V. V. Dobrovitski for inspiring discussions on the suppression of the magnetic field Zeeman effect. This work is supported by National Basic Research Program of China Grant Nos. 2013CB922003 and 2014CB921401, National Natural Science Foundation of China under Grant Nos. 11275139, 11434011, 11121403, and 91421102, National Science Foundation (US) (Grant No. 1208828), NSAF Grant No. U1330201, and the Fundamental Research Funds for the Central Universities.

\appendix
\section{Dynamical decoupling}
\label{appd:dd}

For a spin-1 condensate in a magnetic field, $H = \omega_z S_z + V$, where $\omega_z$ is the Zeeman splitting and $V$ is the weak two-boday interaction (including the spin exchange and dipolar interaction). Suppose the control pulse $\pi_x$ is an instantaneous (hard) pulse and the delay between two adjacent pulses is $\tau$. Note that the $V$ is unaffected but $S_z$ becomes $-S_z$ by the $\pi_x$ pulses. Let us consider the evolution after 2 pulses~\cite{Slichter92},
$$
U(t=2\tau) = U(\tau)U_{\pi_x}U(\tau)U_{\pi_x}.
$$
Since $U_{\pi_x} = \exp(-i\pi S_x) = U_{\pi_x}^\dag $ and $U(\tau) = \exp[-i\tau(\omega_z S_z + V)] \approx \exp(-i\tau \omega_z S_z)\,\exp(-i\tau V)$ for small $\tau$ (i.e., $||H||\tau \ll 1$), we find that
\begin{eqnarray}
U(t=2\tau) &=& U(\tau)U_{\pi_x}U(\tau)U_{\pi_x}^\dag \nonumber \\
&\approx & U(\tau) \exp(-i\tau V) U_{\pi_x}\exp(-i\tau \omega_z S_z) U_{\pi_x}^\dag \nonumber \\
&\approx & \exp(-i2\tau V).
\end{eqnarray}

After periodically applying $2K$ pulses, the evolution operator is
\begin{eqnarray}
U(t=2K\tau) &=& [U(\tau)U_{\pi_x}]^{2K} \nonumber \\
&\approx & \exp(-it V)
\end{eqnarray}
to the leading order of $\tau$. Clearly, the effect of the magnetic field is removed if $\tau\rightarrow 0$.

A possible way to realize experimentally the hard pulse is by applying a square pulse of constant $B_1$, e.g. 10 mG, along $x$ direction within a pulse width $w \sim 10^{-4}$ s. The pulse width is much smaller than the typical pulse delay $\tau \sim 10^{-2}$ s and the dipolar dynamics time scale $\sim 10$ s~\cite{diplifetime}.

%\bibliography{e:/doc/mybib/bec, e:/doc/mybib/sd}

\begin{thebibliography}{41}
\bibitem [{ntd()}]{ntdsn}%
  \BibitemOpen
  \href@noop {} {}\bibinfo {note} {In general, $d_{s,n} = 0$ ($d_n=0$) if the
  mode function has a spherical (an axial) symmetry. For a Gaussian mode
  function with characteristic lengths $q_{x,y,z}$ in $x,y,z$ directions,
  $d_s>0$ ($<0$) if $q_{x,y}<q_z$ ($q_{x,y}>q_z$), and $d_n >0$ ($<0$) if
  $q_x<q_y$ ($q_x>q_y$).}\BibitemShut {Stop}%
\bibitem [{dip()}]{diplifetime}%
  \BibitemOpen
  \href@noop {} {}\bibinfo {note} {The lifetime of a spin-1 BEC longer than 15
  seconds has been observed in experimental groups of M. S. Chapman (Georgia
  Institute of Technology) and Ruquan Wang (Institute of Physics, Chinese
  Academy of Science) (private communications).}\BibitemShut {Stop}%
\end{thebibliography}
%merlin.mbs 2010-03-15 4.21a (PWD, AO, DPC)

%%%%%%%%%%%%%%%%%%%%%%%%%%%%%%%%%%%%%%%%%%%%%%%%%%%%%%%%%%

\end{document}